\documentstyle[12pt,epsfig]{article}   
      
\title{A joint analysis of the $S$--wave in the  $\pi^{+}\pi^{-}$ and $\pi^{0}\pi^{0}$   
 data}   
\author{R. Kami\'nski, L. Le\'sniak and K. Rybicki \\    
Henryk Niewodnicza\'nski Institute of Nuclear Physics,\\    
PL 31-342 Krak\'ow, Poland}   
    
\newcommand{\be}{\begin{equation}}   
\newcommand{\ee}{\end{equation}}   
\newcommand{\ba}{\begin{eqnarray}}   
\newcommand{\ea}{\end{eqnarray}}   
\newcommand{\pipi}{$\pi\pi$ }   
\newcommand{\pipm}{$\pi^{+}\pi^{-}$}   
\newcommand{\pizz}{$\pi^{0}\pi^{0}$}   
\newcommand{\KK}{$K\overline{K}$ }   
\newcommand{\fo}{$f_0(980)$ }   
\newcommand{\ro}{$\rho(770)$ }   
\newcommand{\fd}{$f_2(1270)$ }   
\newcommand{\rf}{$\rho_3(1690)$ }   
   
\newcommand{\aone}{$a_1$ }

\newcommand{\mpp}{$m_{\pi\pi}$ }   
\newcommand{\sw}{$S$--wave }

\newcommand{\downp}{"down--flat" }   
\newcommand{\downs}{"down--steep" }   
\newcommand{\upp}{"up--flat" }   
\newcommand{\ups}{"up--steep" }

\begin{document}   
   
\maketitle   
   
\vspace{-7.6cm}

\hspace{10cm} IFJ-1897/PH/2002

\vspace{6.4cm}

\begin{abstract}   
    
We use our former results on \pipm ~\sw obtained in a nearly    
assumption-free way from the  17.2 GeV/c data to predict the \pizz   
~~$S$--wave. The predictions are compared with the recent results of the E852   
experiment at 18.3 GeV/c. A good agreement is found for only one (the \downp)   
solution while the second one (the \upp) is excluded by the   
~\pizz ~data. Thus the long-standing "up-down" ambiguity has been   
finally resolved in favour of the \sw ~intensity which stays large and nearly   
constant up to the $K\overline{K}$ threshold.    
A joint analysis of both sets of data leads to a reduction of errors for this
 solution.

\end{abstract}   
    
    
\section{Introduction \label{introd}}   
\hspace{0.6cm}  
 Despite of several decades of efforts a spectrum and properties of scalar 
mesons are still not well known. 
This is particularly true in the   
low mass region (below the \KK  threshold), where the whole information comes    
from the \pipi partial wave analysis (PWA). It is interesting to note that    
the best experimental data on the \pipm ~system were collected in the    
seventies. We mean here e.g. the CERN-Munich experiment \cite{grayer}, which    
supplied 3$\times$10$^5$ events of the reaction   
\be   
\pi^-p\rightarrow \pi^+\pi^- n   
\label{reaction1}   
\ee   
at 17.2 GeV/c. It should be stressed that the number of observables provided    
by such experiment is much smaller than the number of real parameters needed    
to describe the partial waves. Consequently, special physical assumptions\footnote   
{Hereafter called the {\it standard} assumptions.} were    
made in most studies [1-4]  
of the \pipm ~system and are still made in all PWA's of the \pizz ~system. It   
has been customary to produce    
results {\it without even mentioning the assumptions} essential for their    
derivation. 
 
These assumptions are generally connected with ignoring of the role  
 of nucleon spin. Also the \aone  exchange amplitude (which is dominantly   
spin-nonflip) was neglected thus assuming the full dominance of the pion   
exchange. It should be stressed that the reality can be   
more complicated, the nucleon spin cannot be {\it a priori} ignored and   
there is a significant pseudovector (the \aone) exchange in addition    
to the dominant pseudoscalar pion exchange.    
This was demonstrated by another important experiment, namely    
that  of the CERN-Cracow-Munich (CCM) collaboration \cite{becker1} which    
yielded $1.2\times10^{6}$ events of the reaction:\\   
\begin{equation}   
\pi^-p_{\uparrow}\rightarrow \pi^+\pi^- n   
\label{reaction_pol}   
\end{equation}   
also at 17.2 GeV/c. The presence of the transversely polarized target provides   
additional observables. It was shown that there is a considerable \aone     
exchange. Combination of the results of both experiments enables   
a quasi-complete PWA with much weaker physical assumptions. The analysis is only   
a quasi-complete one because we do not know the phase between two sets of   
transversity amplitudes. Nevertheless e.g. the {\it full} (containing both    
pseudoscalar and pseudovector exchange as well as both $I=0$ and $I=2$   
components for \sw  and $D$--waves) partial wave intensities could be determined in    
a completely model-independent way. This was done in another CCM    
paper \cite{becker2}. While for dipion mass    
\mpp $=(1000\div1800)~$MeV a unique solution    
was obtained, the up-down ambiguity reappeared below 1~GeV.\\     
\hspace*{6mm}In our first paper \cite{klr} (hereafter called paper I) we   
"have constructed a bridge" between two sets of transversity amplitudes   
for the $S-$wave. This was done by first demanding the phases of the leading    
$P$--, $D$-- and $F$--transversity amplitudes to follow exactly the phases of the    
Breit--Wigner \ro, \fd and \rf resonant amplitudes in the low, medium    
and high mass region, respectively. Then, from the measured phase    
differences between these waves and the \sw we calculated the    
{\it absolute} phases of the \sw  transversity amplitudes\footnote   
{Let us note that such a procedure would be impossible for the low mass   
\pizz ~\sw  even if the reaction (3) was measured on a polarized target.    
There would be no $P-$wave to be used as a handle to determine the absolute    
phase of the \sw.}.    
Finally the pseudoscalar and pseudovector exchange amplitudes in the \sw were    
explicitly determined from the relevant combinations of the \sw     
transversity amplitudes. In order to calculate the $I=0$ components we    
have used the $I=2$ ~\sw as measured by the CERN-Munich \cite{hoogland}   
collaboration. This separation, first ever made, has    
been done using much weaker assumptions\footnote   
{The main assumption was neglecting of a possible influence of the \aone    
exchange in $I=2$ \sw. This assumption was also made in all other studies.} 
than those made in {\it any} earlier    
analysis.  
 
\newpage 
 
The price we paid was a fourfold ambiguity in our pseudoscalar    
exchange \sw  amplitude.  
In addition to the "up-down" ambiguity\footnote{The "down" \sw intensity stays  
high and nearly constant from $m_{\pi\pi}\approx750$ MeV to the \fo while the  
"up" solution  exhibits a resonant-like maximum under the \ro.} of the old    
CCM analysis \cite{becker2} there were ambiguities resulting from adding or    
subtracting the phase difference since the PWA of the CCM group    
yielded only the absolute value of the phase difference.   
Thus we have \downp, \downs, \upp  and \ups  solutions.   
Differences between "flat" and "steep" refer mainly to the behaviour   
of the \sw  phase shifts below the \fo . Above the \fo resonance all the    
solutions are fairly similar. The main difference is that both "steep"    
solutions contain a relatively narrow\footnote   
{Hereafter "relatively narrow" means "with a width close to    
$\Gamma_\rho = 150$ MeV".}    
resonance under the \ro  (like the old "up" solution). Such resonance was    
persistently claimed by Svec \cite{svec}. On the other hand,    
both "flat" solutions indicate a $f_0(500)$ state \cite{kll} with a width    
of about    
$500$ MeV. In particular the "down-flat" solution is essentially similar   
to the old solution of the CERN-Munich group \cite{grayer} obtained with   
the help of the standard assumptions.\\   
\hspace*{6mm}Having arrived at as many as four solutions we   
have immediately started a thorough investigation of them in a hope of   
reducing their number. Already in paper I we have shown the \downs     
solution to violate unitarity while both "flat" solutions easily passed   
this test. The \ups  solution although behaving in a queer way could not be    
immediately excluded.      
Our next paper \cite{klr2} (hereafter called paper II) was    
devoted to a detailed study of this   
suspicious solution. Using the experimental fact that there is no significant    
inelastic channel below the \fo  ~we have imposed the $\eta=1$ inelasticity   
for all mass points in this mass region. For the \ups   solution this led    
to unphysical results in 7 out of 20 mass bins and queer behaviour in the    
remaining mass bins. The effect was even stronger for the \downs ~solution,   
already rejected in paper I. Thus removing both "steep" solutions we were    
left with two "flat" solutions equally well describing the \pipm ~~data.    
The further step demanded an extra input.\\   
\hspace*{6mm}Fortunately such an input has been recently provided by the   
BNL E852 collaboration \cite{E852}, which has collected $8.5\times10^{5}$   
events of the reaction    
\be   
\pi^-p\rightarrow \pi^0\pi^0 n   
\label{reaction2}   
\ee   
at 18.3~GeV/c. A comparison of both "flat" solutions to the \pizz ~~data is   
a topic of the present paper. \\   
\hspace*{6mm}The paper is organized as follows. In Sect. 2 we describe   
the \pipm ~~and \pizz ~~data. In Sect. 3 we relate our \pipm ~   
\sw  amplitudes to the amplitudes describing the \pizz ~~channel, in which the   
role of the isospin 2 component is substantially enhanced in comparison with the   
\pipm ~~channel. In Sect. 4 we compare    
our predictions to the \pizz ~data and use both sets of data to improve    
our knowledge of the \pipi \sw. The results are summarized in Sect. 5.   
   
\section{Description of the \mbox{\boldmath $\pi\pi$} data, first comparison and   
relative normalization\label{descr}}   
   
\hspace{0.6cm}There were several experiments yielding \pizz ~data from   
reaction (3). The highest statistics was provided by the GAMS experiment   
which collected $1.5\times10^{6}$ events \cite{gams38} at 38~GeV/c and    
$6.4\times10^{5}$ events \cite{gams100} at 100~GeV/c. However these high-energy    
experiments were mainly dealing with the high \mpp  region. The NICE   
experiment \cite{nice} at 25~GeV/c, the  KEK E135 experiment \cite{E135}    
and some, even older, experiments were   
 of inferior statistics. In this paper we use the results of the E852    
collaboration \cite{E852} obtained at 18.3~GeV/c ~i.e. very close to the    
CCM momentum of  17.2 GeV/c. Additional (and essential) advantage of these    
results is their availability on the World Wide Web where explicit numbers    
are given. These are intensities of partial waves in    
$\Delta m_{\pi\pi}=40$~MeV bins for    
four intervals of the four-momentum transfer $t$ between --0.01 GeV/c$^{2}$   
and --1.5 GeV/c$^{2}$.\\   
\hspace*{0.6cm}For a comparison with our results we have    
merged two $|t|$ bins i.e. $(0.01\div0.10)$ ~\mbox{GeV$^{2}$/c$^{2}$} and 
$(0.10\div0.20)~$GeV$^{2}$/c$^{2}$ into a single one i.e.    
$(0.01\div0.20)~$GeV$^{2}$/c$^{2}$.   
This was done by adding the values and their errors linearly and in quadrature,   
respectively. This binning is exactly equal to that in the original CCM    
paper \cite{becker2} while in papers I and II we used    
$\Delta m_{\pi\pi}=20$~MeV bins   
and $|t|=~(0.005\div0.200)$GeV$^{2}$/c$^{2}$. \\   
\hspace*{0.6cm}It should be noted that both experiments have certain problems    
at low $m_{\pi\pi}$.   
The spark chamber CM experiment \cite{grayer} had low and badly controlled   
acceptance for two close tracks; thus it was decided not to publish data   
below $600$~MeV. The E852 data \cite{E852} start right from the threshold    
but they are somewhat disturbed by some $K^{0}_{s}\rightarrow\pi^{0}\pi^{0}$    
decays which escaped rejection. The upper \mpp limit of 1800~MeV (1600~MeV)   
for $|t|>0.01~$GeV$^{2}$/c$^{2}$ ($|t|>0.005~$GeV$^{2}$/c$^{2}$) in the CCM    
data is of lesser   
importance for our study of ambiguities present mainly below the \fo .     
However a warning should be added concerning the \pizz  ~data at   
\mpp $=(1400\div1600)$~MeV. An inspection of Fig. 5D and Fig. 7D of    
ref. \cite{E852} shows a substantial $G~(J=4)$ wave below $1600$~MeV, which   
seems to be very unlikely at such a low mass. Consequently the \pizz~  \sw  is    
probably underestimated in this mass region.\\   
\hspace*{0.6cm}Unfortunately the E852 collaboration has provided only the   
acceptance-corrected numbers of events  and {\it not the absolute cross    
section} which is available for the CCM data. Thus we have determined the    
normalization to our data using the strongest $D_{0}$ wave, generally well   
fixed by the $t^{4}_{0}$ moment and saturated by the $f_{2}(1270)$.   
The fit in the range \mpp $=(1000\div1600)$~MeV yields the normalization    
factor $F_{d}=(8.58\pm0.16)\times10^{-6}\mu b$/event in $20$~MeV bins.   
This normalization factor was then applied to other waves.   
The results are shown in Fig. 1.\\   
\begin{figure}[htbp]\centering   
\mbox{\epsfxsize 13cm\epsfysize 19cm\epsfbox{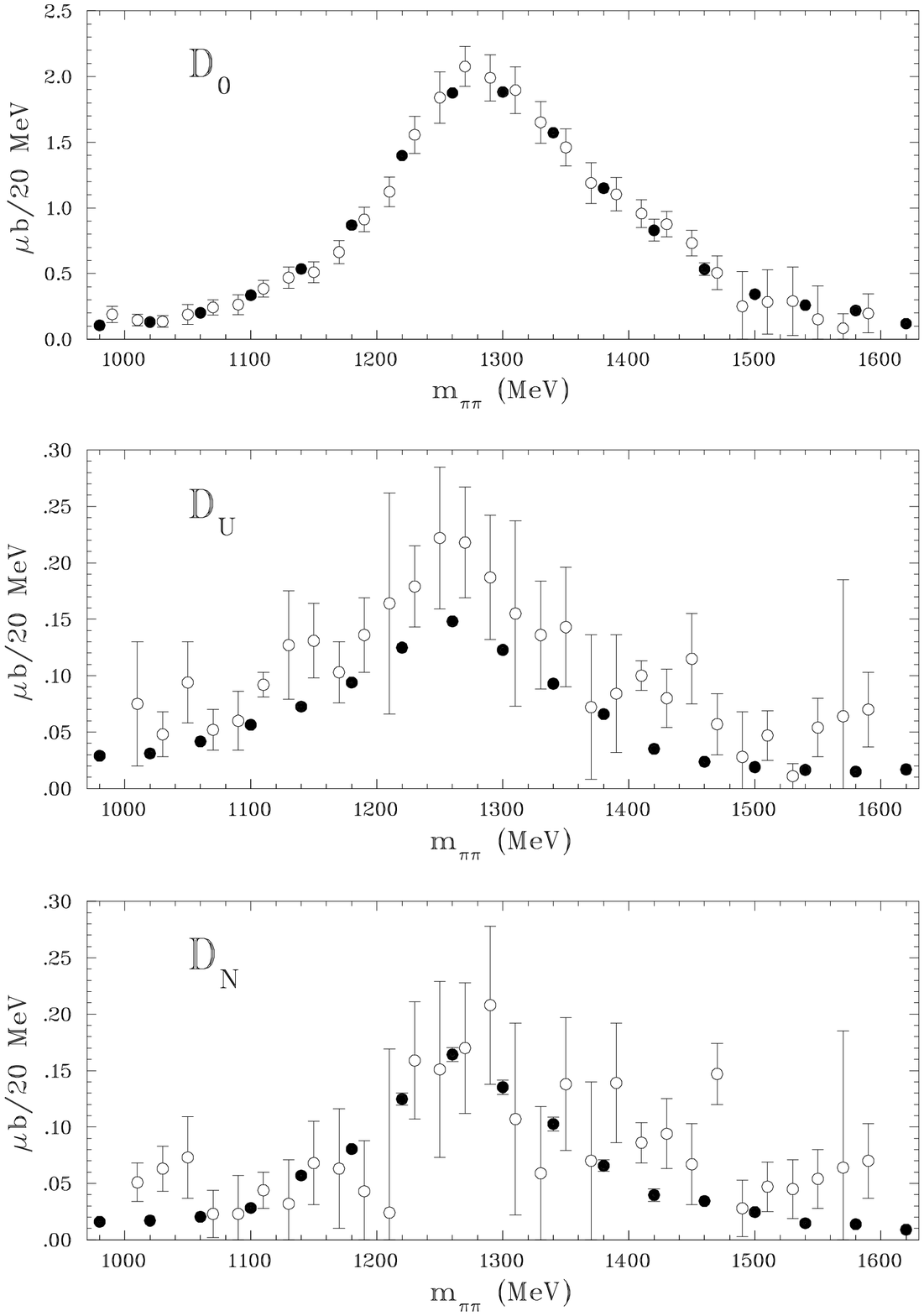}}\\   
{Fig. 1: Intensity of $D$--waves for the \pizz ~data    
(full circles) and the \pipm ~data (open circles)}    
\end{figure}   
\hspace*{0.6cm}An inspection of Fig.1 shows:\\   
-- a good agreement {\it in the shape} of the $D_{0}$ wave;    
this shows that this wave can be used for normalization,\\   
-- a systematic difference in the $D_{U}$ wave (unnatural spin parity exchange)    
with virtually all E852 points below the \pipm ~data,\\   
-- a similar but much less significant effect in the $D_{N}$ wave (natural spin    
parity exchange).\\   
\hspace*{6mm} Now let us concentrate on the $D_{U}$    
wave. We attribute the difference to the assumptions described in the    
introduction, which were used in the PWA for \pizz ~but not for the \pipm    
~data. This effect was already observed in the CCM paper (see Fig. 5 of    
ref. \cite{becker2}), when the intensity of the $D_{U}$ wave obtained using the    
data from the polarized target experiment was compared to that from the    
analysis \cite{estma} based on the standard assumptions.   
The assumptions led to an underestimation of the $D_{U}$ wave and the    
overestimation of the least-fixed $S$-- wave, both    
the total intensity and that of the  $D_{0}$-- wave being fixed by the    
relevant moments.   
Assuming the same mechanism we have tentatively calculated the excess   
of the \pipm ~$D_{U}$ intensity over the \pizz ~~one and subtracted it   
from the \sw ~intensity in the latter. This leads to a $(17\pm4)\%$     
reduction of the \sw ~in the \pizz ~plot and the normalization factor   
$F_{s}^{'}=(7.13\pm0.35)\times10^{-6}~\mu b$/20 MeV per one event.   
Including the above underestimation of the   
$D_{N}$ wave by about 10 $\%$ yields    
$F_{s}^{''}=(6.44\pm0.52)\times10^{-6}~\mu b$/20 MeV per one event.   
Although this whole procedure looks like a step in the wrong direction    
(generally the    
\pizz ~data are already {\it below} the \pipm ~ones) we shall see later that   
a good agreement in one of the \sw ~solutions can be reached after properly    
handling the isospin $I=2$ component of the \sw ~amplitudes. Since we do not   
know whether $F_{s}^{'}$ or $F_{s}^{''}$ is the proper value we use the    
intermediate value   
$F_{s} = (6.70\pm0.78)\times10^{-6}~\mu b$/20 MeV.   
The error range covers both values including their error bars. The value $F_{s}$   
was used to normalize the E852 data and Fig.2 shows a comparison of the cross    
sections found in paper I for two solutions: \upp and "down-flat". The errors   
of the $\pi^0\pi^0$ intensities were obtained by adding the normalization and  
statistical errors in quadrature.   
\begin{figure}[htbp]\centering   
\mbox{\epsfxsize 13cm\epsfysize 19cm\epsfbox{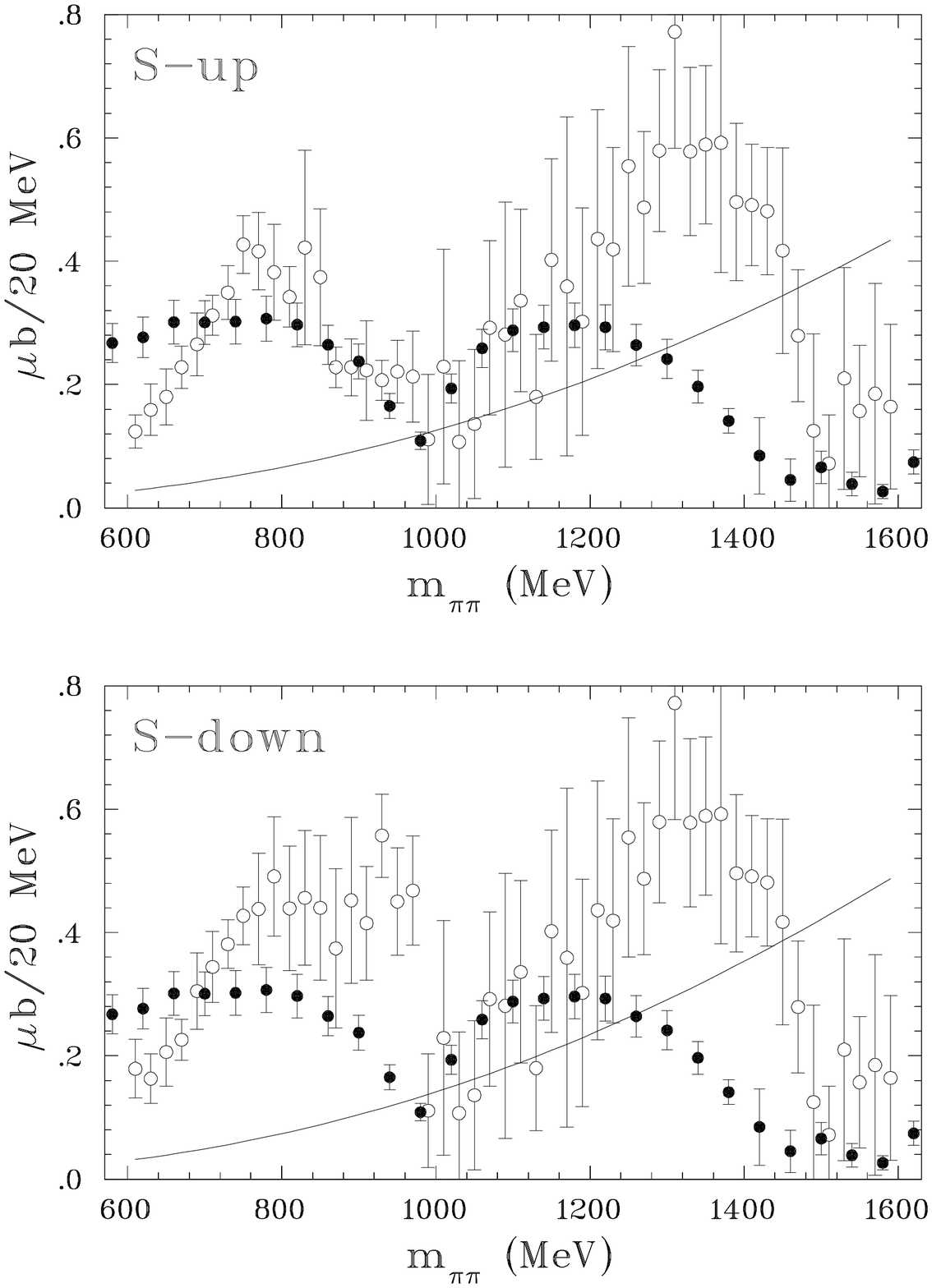}}\\   
{Fig. 2: Intensity of \sw for the \pizz ~data (full circles) and the \pipm    
~data (open circles). Solid line denotes the $I=2$ contribution  
 $2\times~|0.5 \overline{C_\pi U} T_2|^2$, see Eqs. (\ref{g_0bar}),     
(\ref{h_0bar}) and (\ref{I_0}).}    
\end{figure}    
It should be noted that we have repeated the calculation   
of the normalization factor for various mass ranges as well as for the   
CCM binning ($\Delta m_{\pi\pi}=40$~MeV and $|t|>0.01$~GeV$^{2}$/c$^{2}$);   
the results change very~little.   
 
The correctness of the above normalization is supported by    
the following reasoning.    
Since the $\pi\pi$ production mechanism in the $D$--wave is very similar to   
that in the $S$--wave we can estimate the isospin 2    
contribution to the $D$--wave amplitude in the $\pi^+\pi^-$ and $\pi^0\pi^0$    
intensities dominated above 1 GeV by the isospin 0 resonance $f_2(1270)$.    
It was found in Ref. \cite{hoogland} that in the $\pi\pi$ effective mass    
range between 1.25 and 1.50 GeV the $I=2$ $D$--wave phase shift is negative and    
its absolute value does not exceed 4 degrees.   
We have estimated the $D$--wave isospin 2 contribution assuming that   
the $D$--wave $a_1$ exchange amplitude is smaller than the corresponding one    
pion exchange amplitude and that the dependence of the vertex form factors    
on $t$ is the same for the $\pi^+\pi^-$ and $\pi^0\pi^0$ production in both    
S-- and $D$--waves. It turned out that the $D_0$ wave intensities in both    
reactions should be equal within about 2 \%. The assumption about the vertex    
form factors is justified by a very similar shape    
(to about 5 \% below $|t|= 0.4 $ GeV$^2$) of the $t$- distributions in the    
$f_2(1270)$ and $\rho(770)$ mass ranges as shown in Ref. \cite{grayer}.\\   
\hspace*{0.6cm}Let us also comment on the errors being much smaller for   
the E852 results. This is mainly due to high statistics  and to smaller   
numbers of partial waves (e.g. the low mass \sw  saturates the    
\pizz ~intensity while in the \pipm ~data it competes with much stronger    
$P$--wave). On the other hand the E852 errors do not include    
the systematic errors of the standard assumptions inherent in their PWA. These    
assumptions were absent in the CCM analysis. In comparison of the both sets of   
 data the large normalization errors are very important.\\   
   
\section{Calculation of \mbox{\boldmath $S$}--wave intensity for    
\mbox{\boldmath $\pi^{0}\pi^{0}$} channel from our analysis   
of the \mbox{\boldmath $\pi^{+}\pi^{-}$} data \label{calc}}   
   
\hspace*{6mm}In this chapter we shall discuss relations between the \pipi     
production amplitudes in various charge combinations. In addition to the    
reactions (1) and (3) we consider the following reaction:   
\be   
    \pi^+ p \rightarrow \pi^+\pi^+ n    \label{reaction++}   
\ee       
which is an essential source of information about the    
\pipi phase shifts in the scalar isospin $I=2$ \pipi channel \cite{hoogland}.    
   
At high energies of incoming pions and at small momentum transfers to the    
target the above three reactions are dominated by the one pion exchange    
mechanism. However, as shown in \cite{becker2} and in I the pseudovector    
$a_1(1260)$ exchange, although smaller than the one pion exchange, is also    
present and cannot be neglected. Let us denote by    
$T(\pi\pi \rightarrow \pi\pi)$ the off-shell elastic or charge exchange \pipi    
amplitudes and by $t(a_1 \pi \rightarrow \pi\pi)$ similar amplitudes in which   
the exchanged $a_1$ is off-shell. The coupling of the exchanged pion to the    
target proton and neutron is relatively well known. This is not the case for    
the $a_1$ coupling but in paper I we have been able to separate the $a_1$   
exchange amplitudes using the polarized target data of   
reaction (2). Now we shall use those amplitudes and assuming   
the isospin symmetries of the T and t amplitudes we shall write down the    
amplitudes corresponding to the $\pi^0\pi^0$ production. Using the isospin    
symmetry \cite{mms} we can write:   
\be   
T(\pi^+\pi^- \rightarrow \pi^+\pi^-)= {1 \over 3} T_0 + {1 \over 2} T_1 +    
{1 \over 6} T_2,                                      \label{p+-}                \ee   
   
\be                                                                     
T(\pi^+\pi^- \rightarrow \pi^0\pi^0)= - {1 \over 3} T_0 + {1 \over 3} T_2,    
                                                       \label{p00}   
\ee   
   
\be                                                                     
T(\pi^+\pi^+ \rightarrow \pi^+\pi^+)= T_2,                                                                                              \label{p++}     
\ee                                                             
where $T_j~(j=0,1,2)$ are the \pipi isospin ($I=0,1,2$) amplitudes.    
They are normalized to Argand's form:   
   
\be    
T_j=\frac{\eta_j e^{2i\delta_j}-1}{2~i},              \label{normT}   
\ee   
where $\eta_j$ are inelasticities and $\delta_j$ are the channel $j$ phase    
shifts. Similarly    
one can write the $a_1 \pi \rightarrow \pi\pi $ amplitudes as follows   
   
\be   
t({a_1}^{+}\pi^- \rightarrow \pi^+\pi^-)=  {1 \over 3} t_0 +{1 \over 2} t_1 +    
{1 \over 6} t_2,                                \label{a+-}   
\ee    
   
\be                                                                    
t({a_1}^{+}\pi^- \rightarrow \pi^0\pi^0)= - {1 \over 3} t_0 + {1 \over 3} t_2   
                                                 \label{a00}   
\ee                                                                     
   
\be   
t({a_1}^{+}\pi^+ \rightarrow \pi^+\pi^+)= t_2,                                                                                   \label{a++}     
\ee    
where $t_j ~(j=0,1,2)$ are again the isospin amplitudes corresponding to the      
production of the \pipi pair by the \aone meson scattered on a pion. Let us    
notice that the   
isotensor amplitudes $T_2$ and $t_2$ enter in different combinations in the    
above equations. After the partial wave projection the isospin $I=1$    
amplitudes will not contribute to the $S$--wave \pipi production so the $T_1$    
and $t_1$ amplitudes can be ignored.   
   
As in paper I we consider two independent $S$--wave \pipi transversity    
amplitudes $g$ and $h$ for the reaction (2) which differ by a proton   
 or neutron spin projection on the axis perpendicular to the production plane    
($g\equiv<n\downarrow|T|p\uparrow>$ and $h\equiv<n\uparrow|T|p\downarrow>$). By    
$g_0$ and $h_0$ we denote the amplitudes corresponding to the reaction    
(\ref{reaction2}) and by $g_2$ and $h_2$ the amplitudes of the reaction    
(\ref{reaction++}).   
   
The amplitudes of the type $g$ or $h$ for those three reactions can be    
expressed as    
sums of two terms proportional to the proper combinations of the amplitudes   
$T_i$ and $t_i$. Thus according to paper I :    
   
\be   
g = C_\pi U ( {1 \over 3} T_0 +{1 \over 6} T_2 ) +    
    C_a V ( {1 \over 3} t_0 +{1 \over 6} t_2 ) ,           \label{g}                                           
\ee   
   
\be                                                        
h = C_\pi U^* ( {1 \over 3} T_0 +{1 \over 6} T_2 ) +    
    C_a V^* ( {1 \over 3} t_0 +{1 \over 6} t_2 ),         \label{h}                                                       
\ee                                                       
where the functions $C_\pi$ and $C_a$ are proportional to the propagators of    
the exchanged $\pi$ and \aone, respectively. The expression for the $C_\pi$   
factor is following:   
   
\be    
C_\pi=-\frac{{m_{\pi\pi}\sqrt{2\cdot\:\frac{g^2}{4\pi}}}}   
{p_{\pi}\sqrt{sq_{\pi}}\,f}\frac{e^{a t}}{m_{\pi}^2-t} , \label{Cpi}                                                           
\ee                                                
where:\\    
$p_{\pi}$ is the incoming $\pi^-$ momentum in the $\pi^-p$ c.m.   
frame,\\   
$m_{\pi}$ is the pion mass,\\   
$s$ is the square of the total energy in the $\pi^-p$ ~c.m. frame, \\   
$q_{\pi}$ is the final pion momentum in the \pipi c.m. frame,\\   
$g^2/4\pi=14.6$ is the pion-nucleon coupling constant,\footnote {This value of  
 the pion-nucleon coupling constant has been used by us in paper I. According to   
new result of Ref. \cite{Ericson} $g^2/4\pi=14.17\pm0.05\pm0.19$. Since $C_\pi$  
depends on the ratio $\sqrt{(g^2/4\pi)}/f$ any change in the value of the   
pion-nucleon coupling constant can be easily incorporated by a proper change of   
the modulus of $f$.}\\   
 and $f$ is the complex correction factor. This phenomenological parameter   
 should account for the averaged $t-$dependence of the pion-nucleon vertex   
  function and an additional small phase of the pion propagator. In paper I   
  this parameter, close to 1, was introduced to satisfy    
the requirement that the average inelasticity coefficient for the $S$ -- wave    
isoscalar \pipi amplitude should be equal to 1 below the \KK threshold. As in    
paper~I we take the slope parameter $a=$ 3.5 GeV$^{-2}$. The complex functions  
$U$ and $V$ which depend on the kinematical variables like \mpp and $t$ are    
defined by Eqs. (16) and (17) of paper~I. The exact functional form of    
$C_a$ (which can be also obtained from paper I) will not be needed in    
further considerations.   
   
Using Eqs. (\ref{p00}) and (\ref{a00}) we can express the amplitudes    
$g_0$ and $h_0$ for the reaction (\ref{reaction2}) in the form analogous to    
Eqs. (\ref{g}) and (\ref{h}):   
   
\be   
g_0 = C_\pi U ( -{1 \over 3} T_0 +{1 \over 3} T_2 ) +    
    C_a V ( -{1 \over 3} t_0 +{1 \over 3} t_2 ) ,               \label{g0}       \ee   
   
\be                                                        
h_0 = C_\pi U^* ( -{1 \over 3} T_0 +{1 \over 3} T_2 ) +    
    C_a V^* ( -{1 \over 3} t_0 +{1 \over 3} t_2 ).         \label{h0}          
\ee                                                       
Finally we give equations for the amplitudes of the reaction (\ref{reaction++}):   
   
\be   
g_2 = C_\pi U  T_2  + C_a V t_2  ,                      \label{g2}            
\ee   
   
\be                                                        
h_2 = C_\pi U^* T_2  + C_a V t_2 .                      \label{h2}         
\ee                                                       
   
The $S$--wave double differential cross section for the reaction (\ref{reaction2})   
 reads:\\   
\be   
 \frac{d^{2}\sigma_0}{dm_{\pi\pi}dt}= |g_0|^2 + |h_0|^2.      \label{sig0}                           \label{i0}   
\ee   
In order to calculate it from (\ref{g0}) and (\ref{h0}) one needs to know    
four complex amplitudes $T_0$, $T_2$, $t_0$ and $t_2$. Even if one exploits    
the knowledge of the    
amplitudes $g$ and $h$ from our analysis of the polarized target data for the   
reaction (\ref{reaction2}) it is impossible to calculate (\ref{sig0}) and find   
 direct   
relations between $g$ and $g_0$ or between $h$ and $h_0$. This would be simpler   
 if the E852 had made an experiment on polarized target. Since   
this was not a case we have to make an additional assumption to simplify   
the amplitudes. We shall neglect the amplitude $t_2$. This assumption was made   
 in past in the approximate determination of the isotensor--scalar amplitude   
$T_2$ \cite{hoogland}. Now we do the same believing that $t_2$ is much   
less important that the amplitude $t_0$ which we do not neglect.\footnote   
{The standard approach was to ignore both amplitudes $t_{0}$ and $t_{2}$.   
 In this analysis we ignore only $t_{2}$ which is a correction to $t_{0}$.}\\   
\hspace*{6mm}Thus the equations (\ref{g}), (\ref{h}), (\ref{g0}) and (\ref{h0})   
can be written as follows:   
\ba   
g = C_\pi U ( {1 \over 3} T_0 +{1 \over 6} T_2 ) +    
    C_a V ( {1 \over 3} t_0 ),           \label{gn}\\   
h = C_\pi U^* ( {1 \over 3} T_0 +{1 \over 6} T_2 ) +    
    C_a V^* ( {1 \over 3} t_0 ),         \label{hn}  \\      
g_0 = C_\pi U ( -{1 \over 3} T_0 +{1 \over 3} T_2 ) +    
    C_a V ( -{1 \over 3} t_0 ),               \label{g0n}\\   
h_0 = C_\pi U^* ( -{1 \over 3} T_0 +{1 \over 3} T_2 ) +    
    C_a V^* ( -{1 \over 3} t_0 ).         \label{h0n}         
\ea   
Now it is easy to relate amplitudes for two reactions in question:   
\be   
g_0 = {1 \over 2} C_\pi U  T_2 - g  ,                      \label{gg}   
\ee   
\be                                                        
h_0 = {1 \over 2} C_\pi U^* T_2 - h .                      \label{hh}  
\ee  
  
From the above equations we infer that apart of the sign the amplitudes    
$g$, $h$ and $g_0$, $h_0$, corresponding to the $\pi^+\pi^-$ and $\pi^0\pi^0$    
production differ mainly by the presence of the term which includes the    
$I=2$ amplitude and enters in different proportions in (\ref{p+-}) and    
(\ref{p00}). This fact has an important consequence on the   
\mpp dependence of the cross section (\ref{sig0}). \\   
\hspace*{6mm}In practical calculations we need amplitude averages over some    
ranges of the momentum transfer squared $t$. In paper I we have calculated    
the amplitudes $\overline{g}$    
and $\overline{h}$ averaged over $-t$ between 0.005 and 0.2 GeV$^2$/c$^2$, so   
over $\Delta t_1=(0.2-0.005)$ GeV$^2$/c$^2$= 0.195 GeV$^2$/c$^2$. Let us suppose  
 that we   
need the amplitudes $\overline{g}_0$ and $\overline{h_0}$ averaged over some    
other range of $-t$ namely $\Delta t$ (for example we shall take    
$\Delta t$=(0.2-0.01) GeV$^2$/c$^2$= 0.19 GeV$^2$/c$^2$). Then one obvious   
renormalization factor of the amplitudes is    
$n=(\Delta t /\Delta t_1)^{1/2}$. This factor will be used for the parts of $g$  
and $h$ corresponding to the $a_1$ exchange where we do not expect a rapid    
variation with $t$ (for details see paper I). The part   
corresponding to the one pion exchange (proportional to the function $C_\pi$)   
is more sensitive to the $t$-variation so we introduce another factor    
$R$ equal to the ratio of the averaged product of $C_\pi U$ over the range   
$\Delta t$ to the averaged product of $C_\pi U$ over the range $\Delta t_1$.   
Then using the definitions of the complex factors $c_1$ and $c_2$   
(Eqs. (32) and (33) of paper I~) one can derive the following relations:   
\be \overline{g_0}={1 \over 2}~ \overline{C_\pi U}~ T_2 +b_g ~\overline{g} +   
b_h ~\overline{h}    
                                                       \label{g_0bar} \ee   
and    
\be \overline{h_0}={1 \over 2}~ \overline{C_\pi U^*}~T_2+{b_h}^* ~\overline{g}+   
{b_g}^*~\overline{h},   
                                                        \label{h_0bar} \ee   
where   
\be  b_g=-(c_1 R+ {c_1}^*) n,                           \label{b_g} \ee   
\be  b_h=c_2 (1 - R) n                                  \label{b_h}   
 \ee   
and $\overline{C_\pi U} $ and $\overline{C_\pi U}^* $ denote the averages of    
the   
products $C_\pi U$ and  $C_\pi U^* $ over the $\Delta t$ range. Using    
(\ref{g_0bar}) and (\ref{h_0bar}) we obtain the effective mass distribution   
integrated over the momentum transfer squared:   
\be \frac{d\sigma_0}{dm_{\pi\pi}} \equiv I_0 = |\overline{g_0}|^2 +   
|\overline{h_0}|^2 .                                       \label{I_0}\ee   
We shall also call it the intensity $I_0$ of the $S$--wave in the reaction   
(\ref{reaction2}). \\   
   
Thus using the analysis of the CCM $\pi^+\pi^-$ data performed in paper I and    
knowing the $I=2$ component we can calculate the $I_0$ intensity for the    
reaction (3) using the phases of the $\overline{g}$ and $\overline{h}$    
amplitudes fixed in paper I.  
Appearance of the new data on the $\pi^0\pi^0$ production   
enables us to relax some assumptions about those phases.   
   
 In paper I the phases   
of the transversity amplitudes $\overline{g}$ and $\overline{h}$ have been    
obtained essentially from the assumed resonant $\rho(770)$, $f_2(1270)$ and    
$\rho_3(1690)$ phases and   
the measured differences of the $S$--wave phase and the above mentioned $P$, $D$   
and $F$ waves. However, in order to account for the $P$ and the $D$ wave phase   
differences in the range of \mpp above 980 MeV, we have introduced an empirical   
additional phase $\Delta$ which was linearly decreasing function of \mpp    
reaching the zero value at 1420 MeV (see Eq. (52) of I). Below 980 MeV where   
there was not enough available partial waves the value   
of $\Delta$ was arbitrarily assumed to be equal to $50.37^{\circ}$. Now    
we can directly calculate the   
values of the $\Delta$ function independently for each \mpp bin using the new    
$\pi^0\pi^0$ results for $I_0$ in addition to the \pipm ~data. In this way one   
can adjust the phases of  $\overline{g}$ and $\overline{h}$ to fit    
{\em simultaneously} the    
$\pi^+\pi^-$ and the $\pi^0\pi^0$ cross sections. Let us note, however, that   
using the $\pi^0\pi^0$ intensity as a single input at a given effective    
$\pi\pi$ mass    
we cannot calculate two independent phases of $\overline{g}$ and $\overline{h}$.   
 Therefore below   
1460 MeV we follow the parametrization of paper I for the phases $\vartheta_g$ and    
$\vartheta_h$ of the amplitudes $\overline{g}$ and $\overline{h}$:   
   
\be   
\vartheta_g=\left\{ \begin{array}{ll}   
\vartheta_g^S-\vartheta_g^P + \theta_{\rho(770)}     
&   
\mbox{for \hspace{.1cm} 600 MeV $\leq m_{\pi\pi} \leq$ 980 MeV}, \\   
\vartheta_g^S-\vartheta_g^D + \theta_{f_2(1270)}+ \Delta   
&   
\mbox{for \hspace{.1cm} 980 MeV $\leq m_{\pi\pi} \leq$ 1460 MeV}, \\   
    
\end{array}   
      \right.   
\label{fazy_g_s}   
\ee   
    
\be   
\vartheta_h=\left\{ \begin{array}{ll}   
\vartheta_h^S-\vartheta_h^P + \theta_{\rho(770)} + \Delta   
&   
\mbox{for \hspace{.1cm} 600 MeV $\leq m_{\pi\pi} \leq$ 980 MeV}, \\   
\vartheta_h^S-\vartheta_h^D + \theta_{f_2(1270)} + \Delta   
&   
\mbox{for \hspace{.1cm} 980 MeV $\leq m_{\pi\pi} \leq$ 1460 MeV}. \\   
    
\end{array}   
\right.   
\label{fazy_h_s}   
\ee   
In the above equations $\theta_{\rho(770)}$ and $\theta_{f_2(1270)}$     
are phases of the resonant amplitudes defined in paper I and the remaining  
differences of the $\vartheta$ values represent the measured quantities in the  
partial wave analysis of the CCM group.   
The numerical values of $\Delta$ have been calculated to {\em reproduce} the 
experimental    
values of $I_0$. In subsequent fits we assume that the errors of phases 
$\vartheta_g$ and $\vartheta_h$ are solely given by the experimental errors
of the differences between phases of the S and P or D waves determined in the 
partial wave analysis of the $\pi^+\pi^-$ data.

Linear interpolation of the 40 MeV bin data of the E852 group   
(including errors) has been applied in order to obtain the intermediate values    
corresponding to the 20 MeV \mpp bins of the CERN-Cracow-Munich group.   
Following the arguments given in Section 2 the effective mass 1460 MeV is the    
maximum value at which we have used the E852 data.    
   
\section{A joint analysis of the \mbox{\boldmath $\pi^{0}\pi^{0}$}   
and \mbox{\boldmath $\pi^{+}\pi^{-}$} data\label{comp}}   
\begin{figure}[htbp]\centering   
\mbox{\epsfxsize 13cm\epsfysize 19cm\epsfbox{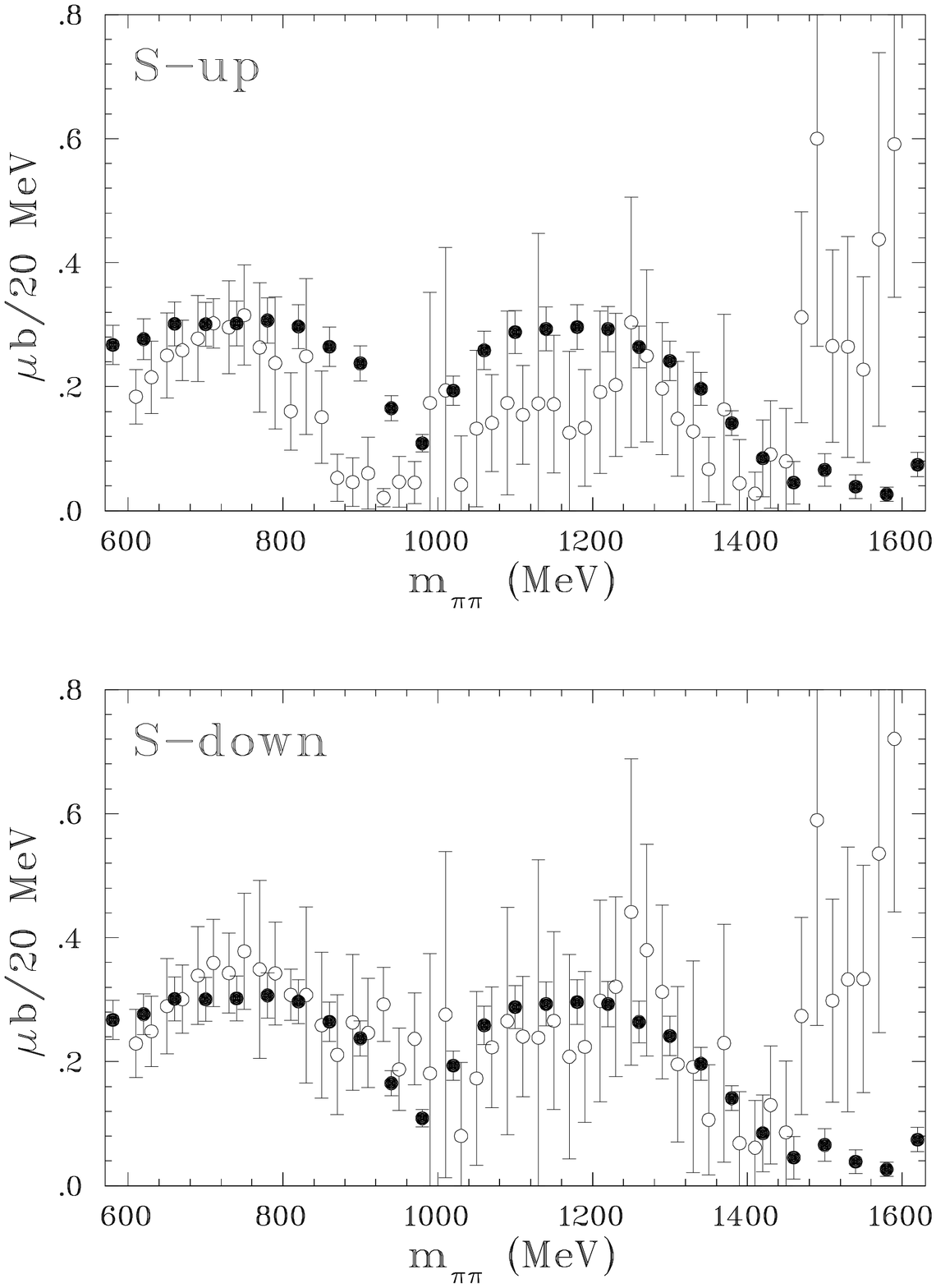}}\\   
{Fig. 3:    
A comparison of the $I_{0}$ determined from the \pipm ~data (open circles)   
and the \sw ~intensity from the E852 \pizz ~data (full circles)   
 normalized with $F_{s}=6.7\times10^{-6}~\mu b$/20 MeV }    
\end{figure}   
\hspace*{0.6cm}Fig. 3 shows the comparison of the \pizz ~intensity $I_0$ from   
eq. (\ref{I_0})   
to the E852 data using the normalization factor $F_s$ determined in Sect. 2.   
Comparing Fig. 2 and Fig. 3 we immediately see an essential role of the   
$I=2$ component. Now it is the "down-flat" solution which reasonably agrees   
with the E852 data apart from the region above $1460$~MeV, for which we   
suspect an underestimation of the \sw  in the E852 data (see~Sect. 2).   
On the other hand the "up-flat" points are too low, especially in the    
region of \mpp $=(870\div970)$~MeV. We have fitted the E852 data to both    
\pipm ~solutions   
with the normalization factor $f_{s}$ as a free parameter.    
This has been done in two mass ranges:\\   
- up to $980$~MeV where the solutions are most different,\\   
- up to $1460$~MeV to avoid the uncertain mass region in the E852 data    
(see Sect. 2). 
   
\begin{table}[ht]   
\begin{center}   
Table 1 \\   
\caption{Results of the fits to the \pizz ~data }   
\vspace*{0.7ex}   
\begin{tabular}{||c||c|c|c||} \hline \hline   
Solution &  Mass range (MeV) & $\chi^{2}/NDF$ & $f_{s}$ (units $10^{-6}\mu$b)\\   
\hline \hline   
"down-flat" &   $600\div980$   &     8.4/18    &     $7.11\pm0.42$    \\   
"down-flat" &   $600\div1460$  &    16.6/42    &     $6.82\pm0.37$    \\   
"up-flat"   &   $600\div980$   &    39.2/18    &     $4.33\pm0.33$    \\   
"up-flat"   &   $600\div1460$  &    45.9/42    &     $4.16\pm0.29$    \\   
\hline \hline   
\end{tabular}   
\end{center}   
\end{table}   

The results are shown in Table 1.  
While the fit is very good for the "down-flat" solution, the   
one for the "up-flat" version shows the strong disagreement. In the low mass    
range there is about $3\sigma$ deviation\footnote{Hereafter $\sigma$ denotes    
a standard deviation.} with the normalization coefficient    
$f_{s}$ being even more significantly ($3.4\sigma$) different from the    
$F^{''}_{s}=(6.44\pm0.52)\times10^{-6}~\mu b/20 $MeV which is    
the lowest possible normalization factor (see Sect. 2). In the whole mass    
range the fit is good but demands the normalization   
factor $f_{s}$ which is $3.8\sigma$ below the corresponding $F^{''}_{s}$.   
If we had not corrected for the influence of the standard assumptions and    
used the $F_{d}$ normalization factor   
then the \pizz ~points would go up by approximately 20$\%$. This could be still   
(although barely) reconciled with the \downp ~solution but the deviation from   
the \upp ~solution would be even more significant. \\   
\hspace*{0.6cm}Let us stress that the fits to the \downp  solution yield    
the normalization factor  $f_{s}$ between    
$F_{s}=(7.13\pm0.35)\times10^{-6}~\mu b$/20 MeV and   
$F_{s}^{''}=(6.44\pm0.52)\times10^{-6}~\mu b$/20 MeV in good consistency   
with both. In particular the fit in the $(600\div1460)$ MeV mass range yields   
 both value   
and error of $f_s$ very close to $F_s=(6.70\pm0.78)\times10^{-6}~\mu b$/20 MeV   
from Sect. 2.\footnote{   
This value of $f_{s}$ means raising the E852 $D_{U}$ wave exactly to the    
level of the \pipm ~data and a small increase of the E852 $D_{N}$ wave    
- see Fig. 1.}   
This confirms that we have applied proper correction to the    
$D_{U}$ and $S$--waves. Now all the $D$--waves (where the $I=2$ component is   
negligible) nicely agree between two experiments   
while the \sw for E852 data is entirely consistent with the $I_0$ for the    
"down-flat" solution of paper I.\\   
\hspace*{0.6cm}Before concluding on a final rejection of the \upp ~solution   
we have critically examined the assumptions introduced in paper I and repeated   
here. The aim of the exercise is an attempt to modify some of the assumptions in such    
a way that the \upp ~solution is saved and we reproduce consistently both \pizz~   
and \pipm ~\mpp distributions. An obvious candidate is our    
assumption concerning the $\Delta$ (see eq. (\ref{fazy_g_s}) and     
eq. (\ref{fazy_h_s})) - phase difference appearing in the \sw ~$g$ and $h$    
amplitudes. \\   
\hspace*{0.6cm}In paper I we dealt with only four observables from the CCM data   
in each mass bin. These are:\\   
$I=|g|^2+|h|^2$ - intensity of the \pipm ~\sw ,\\   
$|g|/|h|$ - ratio of the moduli of transversity amplitudes of the above wave,\\   
two phase differences (for $g$ and $h$) between \sw  and the    
higher wave which is resonating in the region in question. 
   
These four observables were used to determine four real parameters (two   
complex amplitudes) so we had to make some assumptions on $\Delta$, which    
enters the expressions for the last two observables. Our ansatz assumed    
the $\Delta$ for the \sw ~to follow the behaviour of the higher partial   
waves (P and D) for \mpp~ $>980$ MeV. However for lower mass the $D$--wave is too    
weak to serve as a reference wave and   
we have arbitrarily assumed  $\Delta$ to be constant and equal to that   
at \mpp $=990$ MeV. As shown above it is in the low mass region that the   
\upp ~solution disagrees with the E852 data. Therefore we check:\\ 
-- whether one can find the values of $\Delta$ providing an agreement of the    
E852 data with the \upp ~solution as well as with the "down-flat" solution,\\   
-- whether such set of values leads to reasonable behaviour of main parameters,   
namely the inelasticity $\eta$ and the phase shift $\delta$. 
   
This was done by using the    
intensity $I_{0}$  of the \pizz ~\sw from the E852 data in addition to   
the observables listed above. A minor difference with respect to paper I    
is using the actual resonance parameters from the last Review of Particle    
Physics edition \cite{RPP}. Also the range parameter for the    
$\rho(770)$ meson width was chosen equal to 5 GeV$^{-1}$ for both solutions    
"down" and "up" (the Breit-Wigner parametrization of the resonant amplitudes   
is written in~paper~I).    
 
We have critically reexamined all the solutions of partial wave analysis done by  
the CCM group.  
In case of more than one solution for a given ("up" or "down") branch we have  
consequently used the continuity criterion.  
Since these solutions had overlapping errors, the changes are small. 
At five points of \mpp (at 630 MeV for the "down-flat" solution and at 990,   
1130, 1250 and 1330 MeV for the solutions "down-flat" and "up-flat") new values of  
the CKM data were chosen.  
However, those differences in data have changed the values of $\delta$ and  
$\eta$ by much less than their errors calculated in paper I.
 
 \begin{figure}[htbp]\centering   
\mbox{\epsfxsize 13cm\epsfysize 19cm\epsfbox{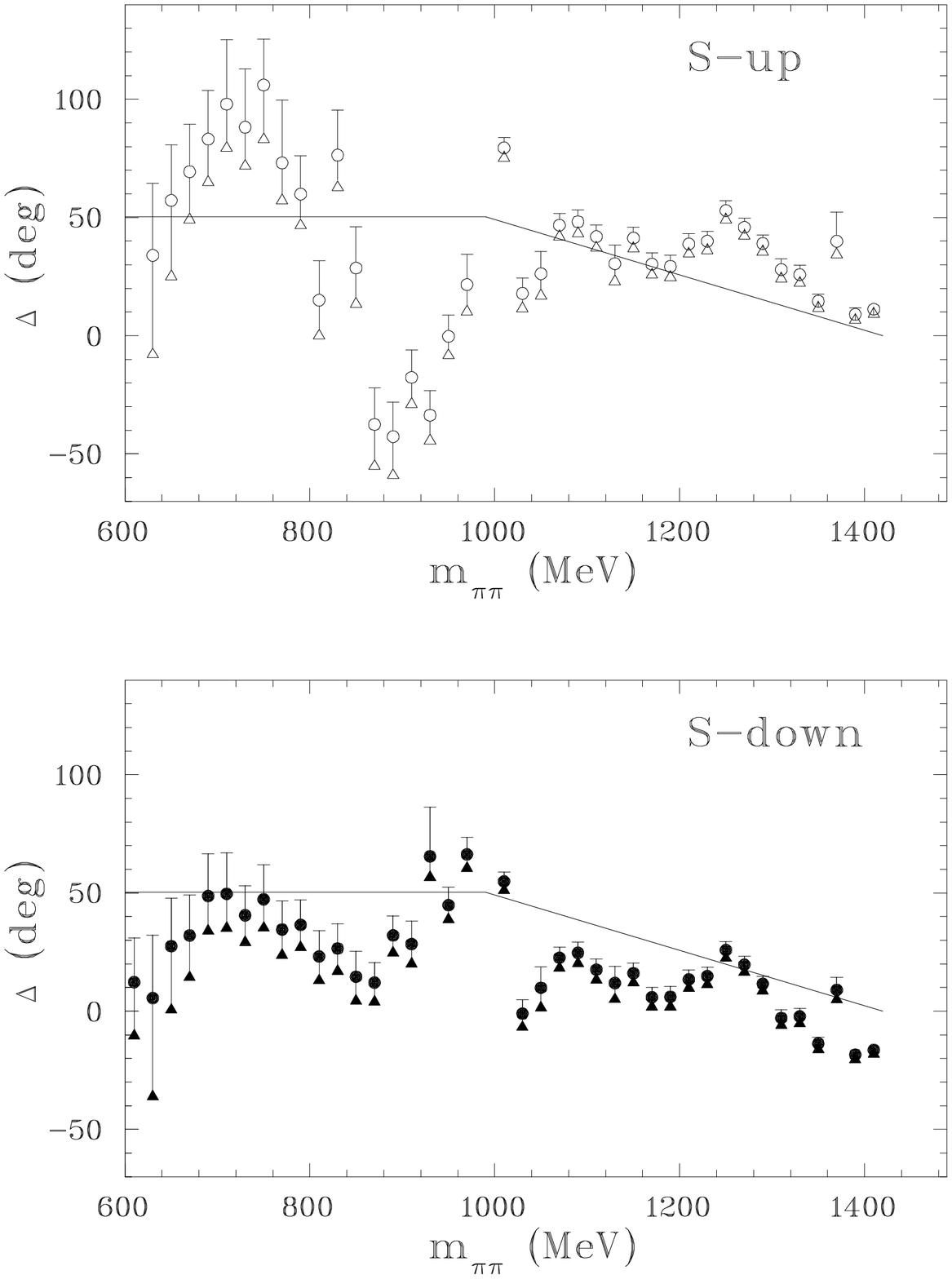}}\\   
{Fig. 4:    
A comparison of the parameter $\Delta$ (circles)    
 with the empirical function of paper I (continuous line). Triangles correspond  
to the maximal values of the $\pi^0\pi^0$ intensity $I_0$ while the upper bars   
denote the values of $\Delta$ related to the lowest values of $I_0$.}   
\end{figure}
   
\begin{figure}[htbp]\centering   
\mbox{\epsfxsize 13cm\epsfysize 19cm\epsfbox{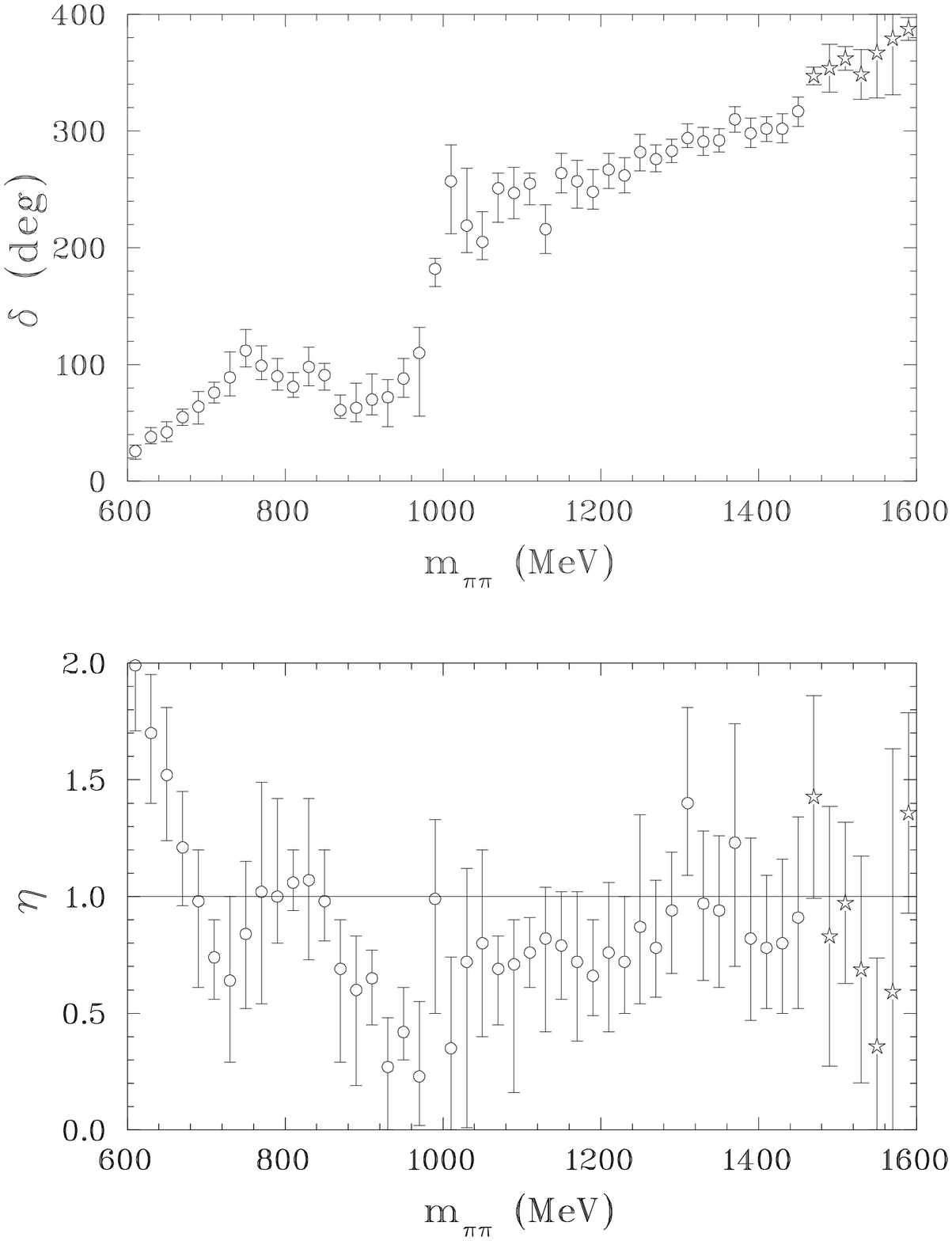}}\\   
{Fig. 5:    
Phase shifts $\delta$  and inelasticities $\eta$ for the \upp ~solution   
(circles). Stars denote data of paper I.}    
\end{figure}
   
\begin{figure}[htbp]\centering   
\mbox{\epsfxsize 13cm\epsfysize 19cm\epsfbox{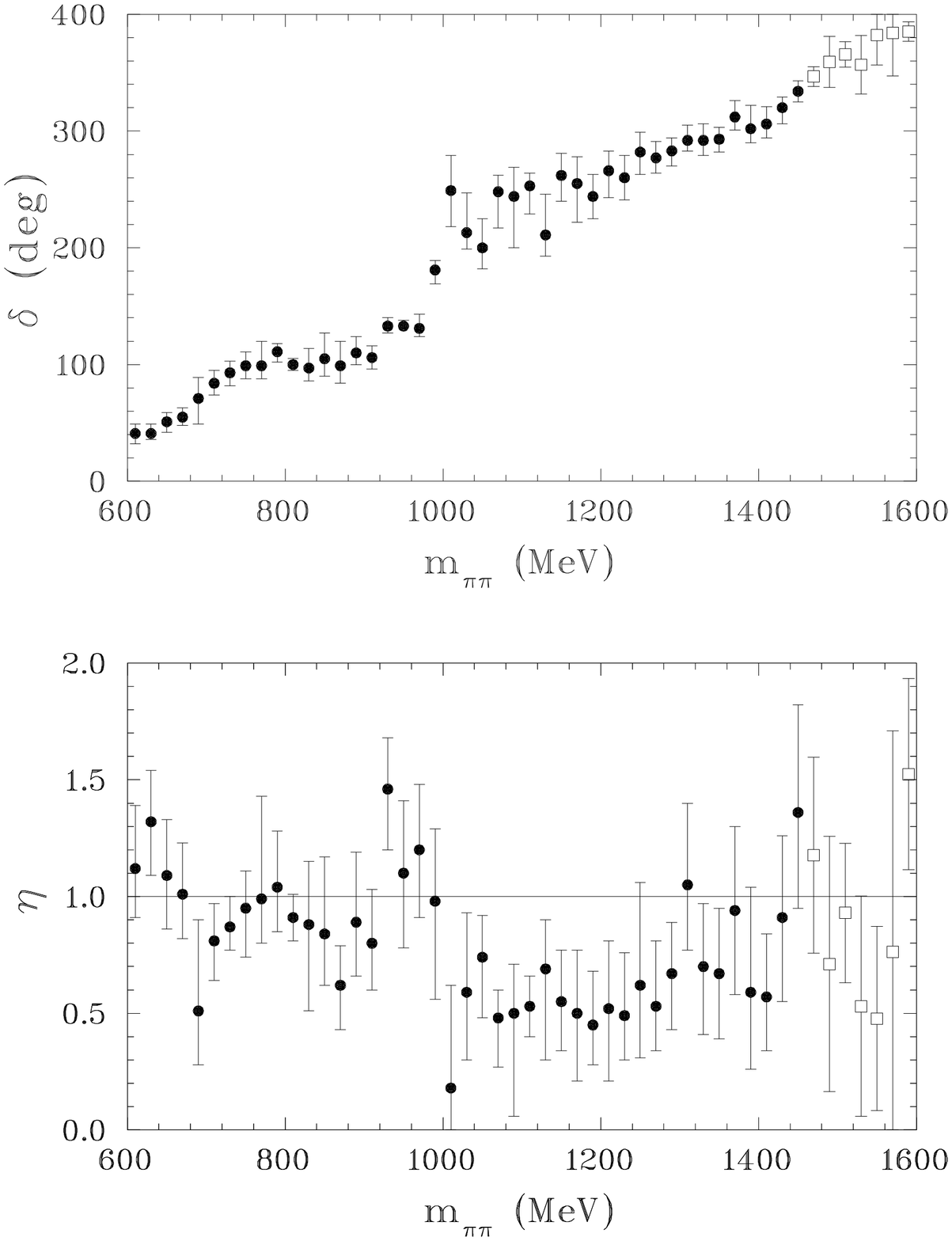}}\\   
{Fig. 6:    
Phase shifts $\delta$  and inelasticities $\eta$ for the \downp ~solution   
(circles). Squares denote data of paper I.}    
\end{figure}   
 
We proceed in the following way: first the moduli of the transversity amplitudes $g$ and $h$  
are calculated from the first two observables, then the \pizz ~data are    
used to determine the values of  $\Delta$ in each mass bin.    
We use the normalization factor $F_s = 6.70\times10^{-6}~\mu b$/20 MeV and  
 calculate the vertex correction factor $f=|f|e^{i\phi}$ (see eq. \ref{Cpi}) by   
minimizing the $(\eta-1)^2$ difference for 19 points of \mpp $<980$ MeV. Now we  
have:\\    
$|f|= 0.814\pm0.019 ,~ \phi= (-5.4 \pm6.8)^{o} $ for the \downp solution compared to    
$|f|= 0.84,~\phi= -17.8^{o}$ in paper I,\\   
$|f|=0.924\pm0.027,~ \phi = (-32.3\pm4.3)^{o} $ for the  \upp solution compared to    
$|f|=0.89,~ \phi =-4.4^{o}$ in paper I.   
 
The resulting   
values of  $\Delta$ are compared in Fig. 4 with the ansatz of paper~I. The   
 variation of $\Delta$ as a function of the effective mass is remarkably rapid  
for the "up" case between 700 and 900 MeV. The values of $\Delta$ drop  
suddenly from about 100$^o$ to negative values as low as $-40^o$. This jump  
is related to a considerable   
difference between the $I_0$ intensities seen in the same mass range in Fig. 3.  
On the other hand for the "down" case the values of $\Delta$ are positive and do  
not vary too much in that range.  
They are also closer to our educated guess on the $\Delta$ made in paper I. 
At 990 MeV the value of   
 $\Delta$ cannot be determined for both solutions since the value of $I_0$ is    
 particularly low at this effective mass (see Fig. 3). The same difficulty is   
 found for two points of \mpp greater than 1420 MeV and smaller than 1460 MeV. 
  Also at 610 MeV for the \upp ~solution the value of $I_0$ is too large to   
 satisfy Eqs. (\ref{g_0bar}), (\ref{h_0bar}) and (\ref{I_0}), so at all the   
 above masses there are no corresponding points in Fig.~4.  
   
  Having new $f$ (the same for each mass bin) and new $\Delta$ (different in   
each bin) we proceed with the calculation of four quantities, above all the    
phase shifts $\delta$ and the inelasticities $\eta$ of the isoscalar 
$\pi\pi\rightarrow\pi\pi$ reaction amplitude, as well as the phases and moduli of the 
$a_1$ exchange amplitude. In each mass bin we demand them to describe five
experimental quantities i.e. the $\pi^0\pi^0$ $S$-wave intensity in addition to
four $\pi^+\pi^-$ observables \footnote {They are: $S$-wave intensity,
$|\overline{g}|/|\overline{h}|$, $\vartheta_g$ and $\vartheta_h$.} used in 
paper I.
The fitting program finds solution and errors in each bin, usually with a very 
small $\chi^2$. The exceptions are 990 MeV, 1430 MeV and 1450 MeV bins for both 
solutions and 610 MeV for the "up" case. Here the $\chi^2$ is larger 
($\sim 0.1$) but reasonable. Above 1460 MeV the fits are not acceptable
as it can be expected from Fig. 3.

The results for the most important parameters   
i.e. the phase shift $\delta$ and the inelasticity $\eta$ are shown in Fig. 5   
and Fig. 6.   
 Unfortunately an    
interesting region around $1500$ MeV is hardly covered by this    
analy\-sis because of the high mass problem in the E852 data. Therefore for   
\mpp larger than 1460 MeV we show the results of paper I. 
  
For the \upp ~solution the inelasticity $\eta$ significantly deviates    
from unity for \mpp $=980$ MeV being too large below $700$ MeV and too small   
above $850$ MeV. This behaviour of $\eta$ is unphysical since there    
are no known inelastic   
processes which can couple to the \pipi channel (see paper II).   
Also the phase shift $\delta$ decreases substantially    
between $750$ MeV and $900$ MeV. The unphysical behaviour of $\eta$ and    
$\delta$ around 900 MeV reflects the large deviation of \pizz ~data from the   
 predictions based on the \pipm ~ "up-flat" solution -- see Fig. 3.   
Thus a change of $\Delta$ or $f$ cannot save the    
\upp solution once the \pizz ~data are included. This conclusion is    
independent of the normalization uncertainties.

 \begin{figure}[htbp]\centering   
\mbox{\epsfxsize 13cm\epsfysize 19cm\epsfbox{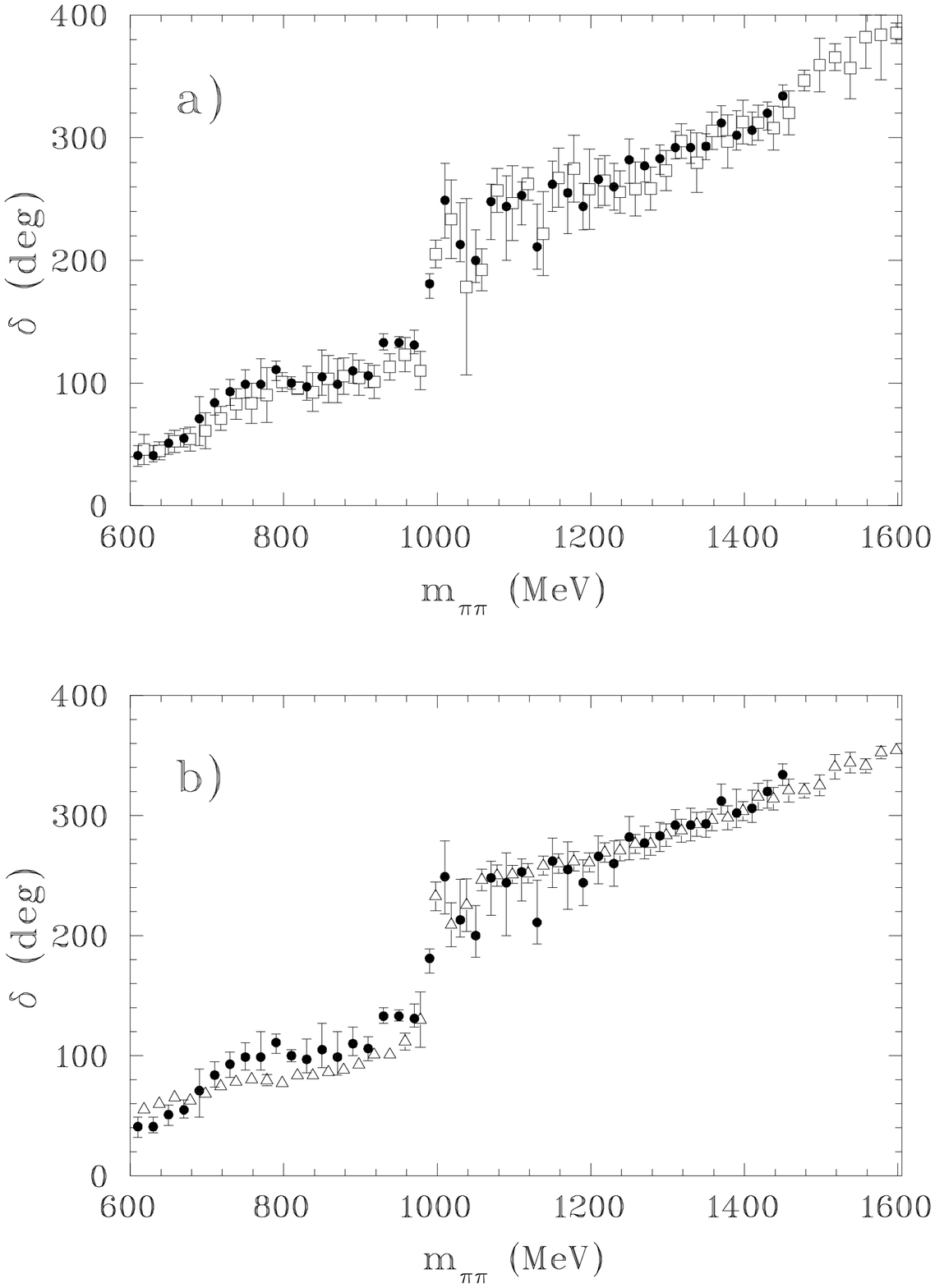}}\\   
{Fig. 7: {\bf a)}   
Comparison of the present phase shifts for the \downp ~solution (circles) with   
 the corresponding phase shifts of paper I (squares - shifted by 8   
 MeV to the right);    
{\bf b)} similar as in a) but triangles - shifted by 8 MeV  
 to the right - denote the results of analysis B in \cite{grayer}. }    
\end{figure}   
  
 \begin{figure}[htt]\centering   
\mbox{\epsfxsize 13cm\epsfysize 8.6cm\epsfbox{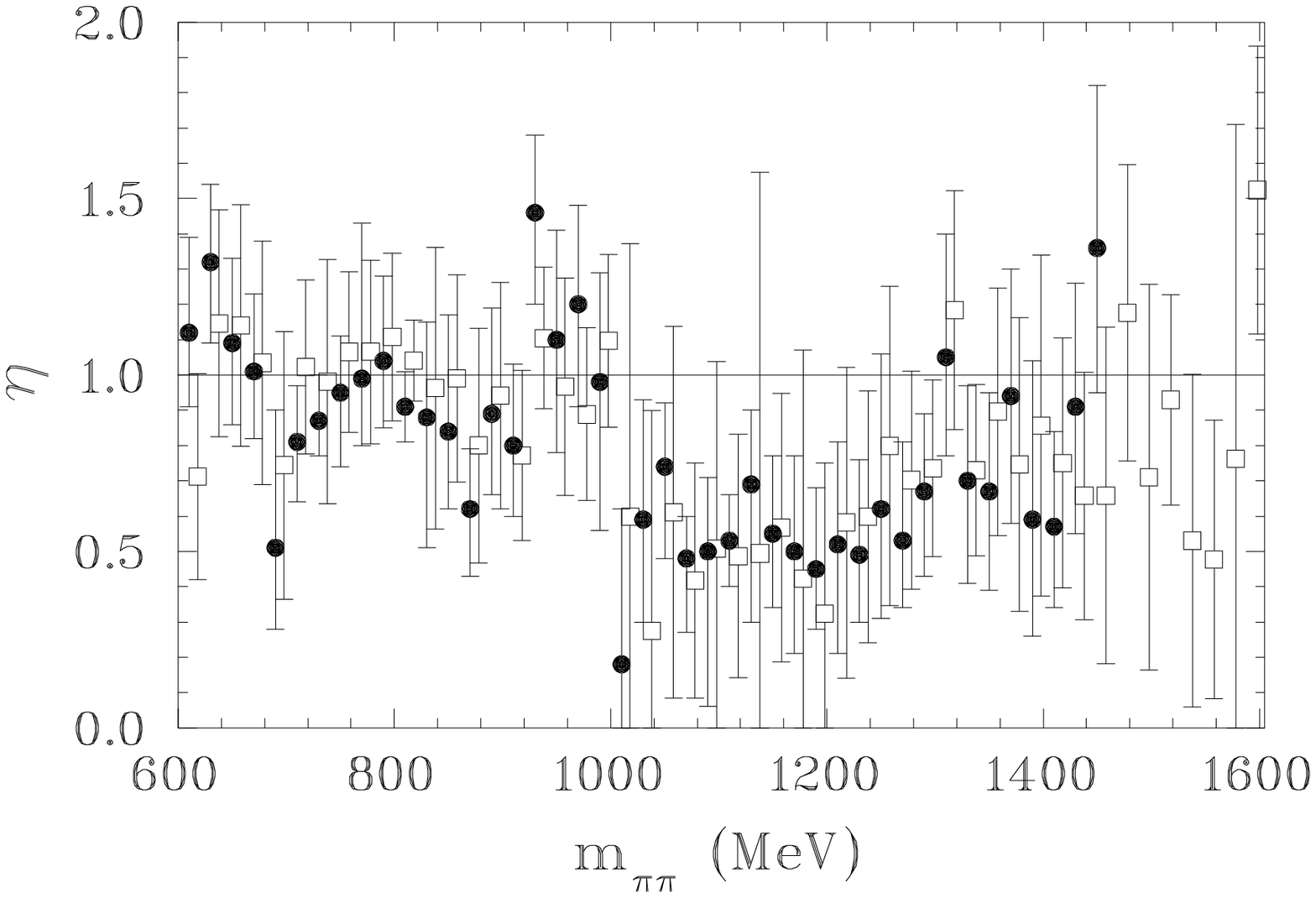}}\\   
{Fig. 8:    
Comparison of the inelasticities for the \downp ~solution (circles)   
with~the~corresponding~ones~from~paper~I~(squares~-~shifted~by~8~MeV~to~the~right)}    
\end{figure}

\begin{figure}[htbp]\centering   
\mbox{\epsfxsize 13cm\epsfysize 17.7cm\epsfbox{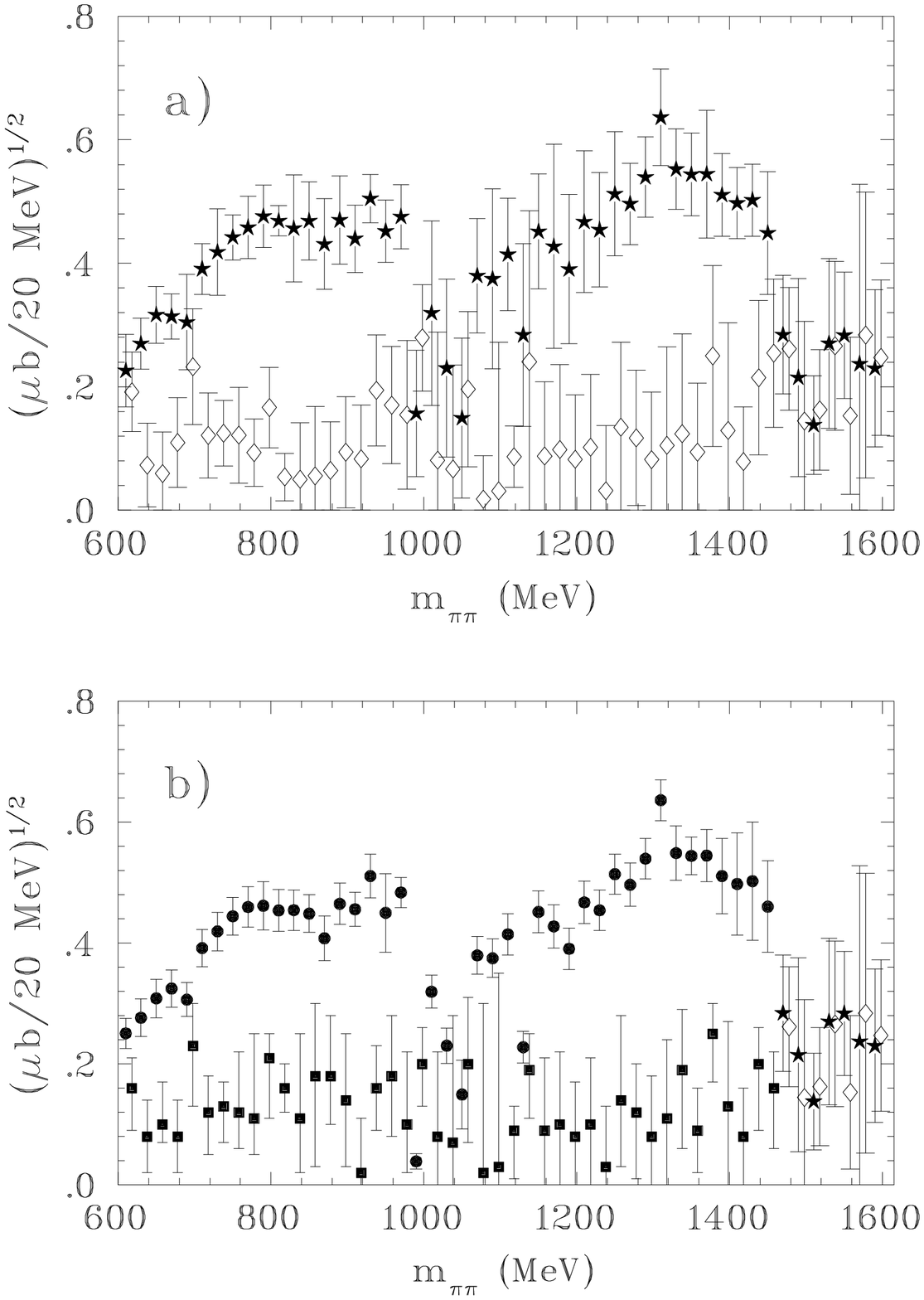}}\\   
{Fig. 9:  {\bf a)}   
Moduli of the pseudoscalar (stars) and pseudovector amplitudes (rhombs   
- shifted by 8 MeV to the right) for the "down-flat" solution calculated using  
 data of paper I;  
{\bf b)} Moduli of the pseudoscalar (circles) and   
 pseudovector amplitudes (squares) for the "down-flat" solution 
of this paper.  
The points above 1460 MeV are the same as in a).}    
\end{figure}   
  
On the other hand the \downp ~solution behaves in a reasonable way.  
Comparing this solution in Fig. 6 with our results from paper I (see Fig. 7a and Fig. 8) we see that 
the additional constraint of the E852 data brought a noticeable reduction of 
errors. In turn comparing the present solution 
with the standard one of Ref. [1] we note some differences in spite of an 
overall similarity (see Fig. 7b). They are clearly seen in the low mass range. The present 
solution starts at lower $\delta$, then crosses the old one around 700 MeV 
and stays above~it~up~to~the rapid change around 990~MeV.

 We have estimated systematical errors due to our particular choice of phases 
$\vartheta_g$ and $\vartheta_h$ in Eqs. (31) and (32). If one determines the 
phase $\vartheta_h$ as in Eq. (32) with $\Delta$ taken from paper I and 
calculates $\vartheta_g$ to reproduce the experimental values of $I_0$ then
the corresponding results for $\delta$ and $\eta$ lie always very well within 
the errors displayed in Fig. 6. Thus we conclude that these systematic errors are
substantially smaller than the errorrs resulting from the joint fits to the
\pipm ~and \pizz  data.

Let us finally comment on the role played by the pseudovector    
(\aone exchange) transversity amplitudes. These amplitudes are significantly   
different from zero. Their moduli are shown in Fig. 9.\footnote{Let us note   
here that in paper I on Fig. 7a) the moduli of the pseudovector amplitudes were   
incorrectly shown due to replacement of the "down-flat" points by the   
 "up-flat" points.} The ratio of the moduli of the pseudovector amplitude to   
 that of pseudoscalar ($\pi$ exchange) amplitudes fluctuates around $0.2\div0.3$ except of the mass regions   
around the \KK threshold and around 1500 MeV where moduli of both amplitudes are comparable.  
In those two regions there exist two scalar resonances $f_0(980)$ and   
$f_0(1500)$. A closer insight into an interplay of the \aone and the $\pi$   
exchange amplitudes in the effective mass distribution reveals particularly   
small values of the interference terms. This result may be the main reason while  
our more sophisticated analysis has produced the solution for the isoscalar   
$S-$wave somehow resembling the old one ~\cite{grayer} outside of the \mpp 
ranges near the $f_0(980)$ and $f_0(1500)$ resonances.  
   
In our opinion these results represent now the best knowledge    
on the \pipi ~\sw in the mass range \mpp $=(600\div1460)$ MeV. They are    
based on the best \pizz~ and \pipm~ data and use the weakest assumptions about
the $\pi\pi$ \mbox{amplitudes}. 
   

\section{Summary}   
\hspace{0.6cm}   
   
We have compared the \pipm~ $S$--wave production data obtained at 17.2 GeV/c by    
the CERN-Cracow-Munich collaboration with the \pizz ~ data recently collected    
by the E852 group. This comparison has been made in two steps.   
In the first step we have used the one pion and $a_1$ exchange model developed    
 in paper I to predict the $S$--wave intensity of the \pizz ~ production at   
18.3 GeV/c.   
Here the previous assumptions about the phases of two independent $g$ and $h$   
helicity amplitudes have been directly taken from paper I.   
Presumably small isotensor part of the $a_1$ exchange amplitude has been   
neglected.   
The proper normalization of the $S$--wave \pizz ~ intensity has been done    
using the strong $D_0$--waves in both experiments.   
The result of the comparison of the calculated \pizz ~ \sw ~intensity with the   
 \pizz ~   
data was satisfactory for the "down-flat" solution but the \pizz ~ intensity   
corresponding to the "up-flat" solution was significantly and systematically    
too   
low in the \mpp range between 800 MeV and 1000 MeV.   
   
In the second step we have varied the phase    
correction $\Delta$ to the $g$ and $h$ amplitudes at each \mpp   
bin trying to fit simultaneously the \pipm ~and \pizz ~\sw~ intensities.   
Then we have recalculated the \pipi scalar-isoscalar phase shifts and   
inelasticities.   
Again for the "down-flat" solution to the \pipm ~data we have obtained a    
reasonable   
behaviour of the inelasticity parameter which,    
as one should expect, was close to unity below the \KK threshold.   
However, the \mpp variation of inelasticity corresponding to the "up-flat"   
solution was too strong.   
Also the "up-flat" phase shifts have an unphysical minimum at 900 MeV.   
This is a clear signal that the "up-flat" solution should be rejected.   
Since the "down-flat" phase shifts and inelasticities behave regularly we    
accept the "down-flat" solution, resulting from the joint analysis of the 
\pipm ~and \pizz ~data as a unique physical one.

   
\addvspace{1cm}   
{\em Acknowledgements}\\   
 The authors are very grateful to Wolfgang   
Ochs and Michael Pennington who independently insisted that such an    
analysis should be done by the Cracow group. This analysis was solely possible   
thanks to the E852 collaboration which stored the numerical values of their   
partial waves on the Web thus allowing free access to them. One of us (L.L.)    
would like to thank Alex Dzierba, Jeffrey Gunter, Adam Szczepaniak and Scott
Teige for valuable discussions.    
   

\end{document}